\documentstyle[12pt]{article}
\newcommand{\be}{\begin{eqnarray}}
\newcommand{\ee}{\end{eqnarray}}
\def\L{\Lambda}
\def\f{\phi}
\def\a{\alpha}
\def\ad{\dot{\alpha}}
\def\pa{\partial}
\def\tr{\mbox{tr}}
\newcommand{\abs}[1]{\left| #1 \right|}

\begin{document}

\begin{titlepage}

\begin{flushright}
PUPT-1760, hep-th/9801121 \\
\end{flushright}

\begin{center}
\vskip3em
{\large\bf Accelerating D-branes} 

\vskip3em
{V.\ Periwal\footnote{E-mail:vipul@phoenix.princeton.edu} and\ R.\ 
von Unge\footnote{E-mail:unge@feynman.princeton.edu}}\\ \vskip .5em 
{\it Department of Physics\\ Princeton University\\ Princeton, NJ 
08544, USA\\}

\vskip2em
\end{center}

\vfill

\begin{abstract}
\noindent Higher derivative terms are computed in the one-loop effective action 
governing the interactions of D3-branes, in two ways: (1) in a 
formalism with $N=2$ supersymmetry, and (2) in 
the standard background field formalism, with only on-shell 
supersymmetry. It is shown that these calculations only agree using 
tree-level equations of motion. The off-shell supersymmetric 
calculation exhibits acceleration terms that appear in terms with 
four derivatives. These may imply disagreement at two-loop order 
between supergravity and Yang-Mills descriptions of D-brane dynamics.

\end{abstract}

\vfill
\end{titlepage}

\section{Introduction}
Witten\cite{witten}
proposed that low-velocity, short-distance interactions of D-branes 
should be described by supersymmetric Yang-Mills(SYM) theories in 
various dimensions, corresponding to the dimensions of the 
world-volumes of the D-branes. The conjecture of Banks, Fischler, Shenker and 
Susskind\cite{bfss} uses this description of D0-brane 
interactions, but interpreted in a
remarkable manner as a non-perturbative definition of M-theory in an 
infinite momentum frame. In this context, the SYM description should 
hold for arbitrary dynamics and at arbitrary distances, with 
agreement with supergravity expected at long distances. Much evidence 
has been accumulated for the validity of this conjecture\cite{banks}.

Some of this evidence relies on phase-shift calculations for 
straight-line trajectories\cite{banks}, finding agreement even at 
two-loop order\cite{bbpt}. Given that finite-time calculations show 
new terms appearing even at one-loop order\cite{oy}, it is of 
interest to consider the  effects of  curved trajectories. 

We do not address Matrix theory\cite{bfss} directly in this paper, 
though this work, combined with the work of Taylor\cite{wati}, is our 
main motivation. We consider
the dynamics of D3-branes as described by $N=4$ 4d SYM theory. We 
calculate higher-derivative terms at one-loop order in two ways.
The first calculation, presented in Sections 2 and 3, is done in a 
formalism that
preserves $N=2$ supersymmetry, and gives $v^{4}/r^{4}$ 
terms at one-loop order, but automatically accompanied by 
$v^{2}a/r^{3}$ and $a^{2}/r^{2}$ terms. The second calculation 
(Section 4), done in a background field formalism with only on-shell 
supersymmetry, exhibits acceleration terms of the form 
$v^{2}a^{2}/r^{6} +$ higher orders. For recent work along
the lines of this calculation, see \cite{tseyt}.
The coefficient of the $v^{4}/r^{4}$ term 
agrees precisely with supergravity in both calculations. We discuss 
this comparison, and further inferences, in Section 5.

\section{N=2 implications}

According to Witten\cite{witten}, the low-velocity short-distance 
dynamics of D3-branes should be described by $N=4$ SYM theory. 
Off-shell $N=4$ supersymmetry in four dimensions is not 
well-understood, so we consider what can be said about the effective 
action in a formalism with only $N=2$ supersymmetry preserved 
off-shell.

The leading term in a momentum expansion of the $N=2$ superspace 
effective action is the imaginary part of a chiral integral of a 
holomorphic function ${\cal F}(W)$ \cite{Fstuff}, where $W$ is the 
$N=2$ gauge superfield strength, whose $N=1$ superfield components 
are the $N=1$ gauge superfield strength $W_{\a}$ and the chiral 
superfield $\f$. The next term in the expansion is a full ($N=2$) 
superspace integral of a real function ${\cal H}(W,\bar{W})$. This 
term was first studied in \cite{mans,mbm}. The function ${\cal H}$ 
must be dimensionless so it is a function of dimensionless 
combinations of $W,\bar{W}$ and some scale $\L$. Furthermore, since 
${\cal F}$ saturates the R-anomaly, ${\cal H}$ must be $U(1)_{R}$ 
invariant. That fixes the form of ${\cal H}$ to be \be
{\cal H} = {\cal H}^{0} + c \left( \ln\frac{W^2}{\L^{2}} + 
g^{0}(W) 				 \right)
\left( \ln\frac{\bar{W}^2}{\L^2} + \bar{g}^0(\bar{W})\right). \ee The 
function ${\cal H}^{0}$ is a function of dimensionless ratios of $W$ 
and $\bar{W}$ and the perturbative one-loop contributions to it have
been calculated \cite{mbm,us}. For a $U(1)$ theory one cannot write 
such a term.

In the case we are interested in, the theory with $N=4$ supersymmetry 
and gauge symmetry $SU(2)$ spontaneously broken to $U(1)$, there is 
no scale and no ${\cal H}^{0}$, and hence the whole term can be 
written as \be
{\cal H} = c \ln\frac{W^{2}}{\L^{2}}\ln\frac{\bar{W^{2}}}{\L^{2}}. 
\label{npform}
\ee
Notice that there is no scale dependence since the terms dependent on 
the scale are purely holomorphic or anti holomorphic and thus are 
killed when integrating over the full $N=2$ superspace measure. In 
\cite{ds} it was argued that since two or higher loop corrections and 
non-perturbative correction would introduce non-trivial scale and 
coupling constant dependence which would violate the symmetries of 
the ${\cal H}$-term, there can be no such corrections. This means 
that (\ref{npform}) is the full perturbative and non-perturbative 
form of this term. The only unknown is the constant $c$. It will be 
calculated in the next section.

There have been several partial checks of the claims in \cite{ds} in 
the literature. The instanton calculations in \cite{dkm,diego} and 
the two loop calculation in \cite{everybody} all seem to support the 
result.

To find the terms relevant to D3-brane scattering in the SYM 
description, we expand ${\cal H}$ in bosonic components assuming it 
has the form (\ref{npform}). The relevant terms are (see also 
\cite{gmm,oz})
\be
\int d^{4}x 4c \left( \frac{\abs{\dot{\f}}^{4}}{\abs{\f}^{4}} 
-\frac{\ddot{\f} \dot{\bar{\f}}^2 }{\f\bar{\f}^2} 
-\frac{\ddot{\bar{\f}}\dot{\f}^2 }{\bar{\f}\f^2} + 2 
\frac{\abs{\ddot{\f}}^2}{\abs{\f}^{2}}
\right).\label{name}
\ee
The complex field $\f$ is to be interpreted as the coordinates of the 
brane in two of the six transverse dimensions (the other two are in 
the scalar components of the hypermultiplet). We may therefore 
rewrite this formula as
\be
4c\frac{v^{4}+2v^2 \left(\vec{r}\cdot\vec{a}\right) 
-4\left(\vec{r}\cdot\vec{v}\right)\left(\vec{a}\cdot\vec{v}\right) 
+2r^{2}a^{2}}{r^{4}}
\label{above}
\ee
Note that this term contains accelerations, as well as the expected 
$v$ dependence, and that this follows purely {}from $N=2$ 
supersymmetry. The only unknown is the constant $c.$ $c$ cannot 
be zero, since that would imply a vanishing potential at 
$v^{4}/r^{4}$ order.

Integrating the terms with acceleration by parts can change the 
coefficient of the $ {v^{4}}/{r^{4}}$ term. However, this also
leads to terms proportional to radial velocities, as well as 
additional acceleration terms. Such changes do not affect the
equations of motion, obviously, so we have written the $v^{4}$ term
with no radial velocity factors. 

\section{N=1 calculation}
We now turn to the problem of calculating the unknown constant $c$ 
{}from the previous section.

The method we use is adapted from that used in 
\cite{mbm,everybody}. The idea is to perform perturbative 
calculations in the bare non-abelian action around an expectation 
value for the chiral field $\f$. If one is interested in the 
effective action for the massless fields only, one can choose all 
external fields to lie in the the same (abelian) direction which 
simplifies the group theory substantially. Furthermore, to calculate 
$c$ it is not enough to keep only $\f$ fields external since for the 
theory we are interested in, there is no contribution to the $N=1$ 
K\"{a}hler potential of $\f$ coming  from ${\cal H}$. Instead we 
focus on another term in the $N=1$ expansion of ${\cal H}$. Following 
\cite{mbm}, we find the $N=1$ expansion
\be
S_{\cal H} = \int d^{4}x d^{4}\theta \left( \ldots + i{\cal 
H}_{A\bar{B}}
\bar{W}^{B\ad}\nabla^{\a}_{\ad}W^{A}_{\a} + \ldots \right), \ee where 
${\cal H}_{A\bar{B}} = \frac{\pa^{2}{\cal H}}{\pa \f^{A}\pa 
\bar{\f}^{B}}$. Note that this is the only term in the $N=1$ 
expansion of ${\cal H}$ that contains two field strengths and no 
derivatives on $\f$.

By performing a calculation with two external vector fields at 
non-zero momentum and an arbitrary number of external $\f$ fields at 
zero momentum we are able to read off the coefficient and we find the 
value
\be
c = \frac{1}{4(4\pi)^2}\ .
\label{XXX}
\ee
This can be checked by comparing to the four gauge boson one-loop
scattering amplitude computed in 4d $N=4$ SYM theory using the 
supersymmetric background field method\cite{book}. 
This result, eq.~\ref{XXX}, disagrees with
calculations performed in harmonic superspace \cite{bko,ketov} where 
the authors find  $c=0$. A detailed presentation of the 
calculation will appear in \cite{nextone}.

\section{The Background Field Method}
We wish now to calculate the effective action at one-loop order in 
the background field method, starting from the dimensional reduction 
of the ten-dimensional SYM action to four dimensions. The point of 
interest for our purposes is that we wish to compute terms with 
derivatives in the effective action. The easiest way of computing 
these is to take advantage of the fact that in background field 
gauge, gauge potentials can only occur in covariant derivatives. 
Thus, if we start from a background field configuration with constant 
gauge potential matrices, we can extract terms in the effective 
action with arbitrary numbers of covariant derivatives. 

The effective action is given by the logarithm of a product of 
determinants at one-loop order. For the $N=1$ 10d SYM theory, with 
fields that realize supersymmetry on-shell, the determinants of 
interest are all of the form $X + F_{mn}J^{mn},$ where $F_{mn}$ is 
the curvature of the gauge potential in the adjoint representation, 
$J^{mn}$ are
the Lorentz generators in appropriate representations (vector, spinor 
or scalar), and
\be X \equiv -\pa^{2} + i(\pa\cdot A + A\cdot \pa) + A\cdot A \ee is a 
Lorentz scalar operator which is a matrix in the (enveloping algebra 
of the) adjoint representation. It is a trivial exercise to obtain 
{}{}from these expressions the dimensionally reduced effective action. 
All derivatives with indices $m,n\in\{4,\ldots9\}$ are set equal to 
zero. We will only need a configuration where $F_{mn}$ is non-zero 
only for indices $m,n$ ranging over four values, say, $m,n\in 
\{0,1,4,5\}.$ Thus, we may as well use the standard ten-dimensional 
generators of the Lorentz algebra in the $F_{mn}J^{mn}$ term. We have 
six  scalars, one vector, two Dirac spinors, and one ghost determinant 
(which exactly cancels two scalar determinants at one-loop order)
in four dimensions, or equivalently, one Majorana-Weyl fermion, one vector
and one ghost determinant in ten dimensions. The 
effective action is
\be \Gamma = \int \sum {(-)^{n}\over n}
\tr_{G}\left(\left((X^{-1}F\right)^{n} \right)\left[{1\over 8}\tr_{F} J^{n} - 
{1\over 2}\tr_{V} J^{n}\right]
\label{below}\ee where $\tr_{G}$ denotes a trace over 
gauge indices, $\tr_{F}$ denotes a trace over Dirac indices, and 
$\tr_{V}$ denotes a trace over vector indices, and all
traces are computed in ten dimensions. The Lorentz indices on 
the product $J^{n}$ are contracted with the Lorentz indices on the 
$F_{mn}$ matrices.

We see immediately that since
\be {1\over 4}\tr_{F} J^{mn}J^{pq} =   \tr_{V} J^{mn}J^{pq} \ee and \be 
{1\over 4}\tr_{F} 
J^{mn}J^{pq}J^{rs} =   \tr_{V} J^{mn}J^{pq} J^{rs} \ee that
if $[A_{4},A_{5}]=0, A_{1}=0,$ there are no terms with fewer than 
four derivatives of $A_{4}$ or $A_{5}.$ The term with $J^{4}$ does 
not cancel between the fermion and vector traces. This is, of course, 
the well-known
fact that there is no renormalization of the coupling constant in 
$N=4$ 4d SYM theory, and that the first term in the effective action 
has four derivatives. Additional derivatives arise in expanding 
$X^{-1}$ assuming $A_{0}$ small. Every term with an $A_{0}$ must 
arise {}from a covariant derivative $D_{0},$ and hence contains all 
necessary information regarding higher partial derivatives. 
Working
out the normalization of the $(\pa_{0}A_{i})^{4}$ term explicitly, we 
find a coefficient $1/(4\pi)^{2}.$   We compare this with
the term $|\dot\phi|^{4}/|\phi|^{4}$ in eq.~\ref{name}, and we 
see that it corresponds to $c = 1/4(4\pi)^{2},$ 
which is precisely the value  given in eq.~\ref{XXX}. This value  
$c=1/4(4\pi)^{2}$
can be checked by 
comparing to the $15/16$ coefficient expected in a dimensional 
reduction to 1 space-time dimension.
Ref.'s~\cite{bko,ketov}  found the  value 
$c=0,$ which disagrees with both our eq.~\ref{XXX}, and our background 
field calculation.

Even though the value  $c=1/4(4\pi)^{2}$ agrees between the background field 
method, and the calculation presented in Sect.'s~2,3, the crucial 
point is that the effective action eq.~\ref{below} does {\it not}
agree with eq.~\ref{name}---we now turn to a detailed examination of 
this discrepancy, in the next section (Sect.~5).

\section{Comparison to supergravity}

We wish to compare the higher derivative terms calculated in these 
two different ways with the interpretation of the 4d SYM action as a 
description of the dynamics of a collection of three-branes in the 
BPS limit, due to Witten\cite{witten}. However, while Witten's 
proposed description was in the limit of small separations, we will 
consider how this description matches up at large separation, 
following Douglas, Kabat, Pouliot and Shenker\cite{dkps}. 

The spacetime metric produced by an extremal three-brane\cite{igor} 
is \be ds^{2}
= f^{-1/2}(-dt^{2} + dx^{\mu}dx^{\mu}) + f^{1/2} dx^{i}dx^{i}, \ee 
where $\mu=1,2,3,$ are indices parallel to the three-brane, $i\in 
\{4,\ldots,9\},$ are transverse indices, and $f(r^{2}\equiv 
x^{i}x^{i} ) \equiv 1+R^{4}/r^{4}.$ The 4-form potential arising {}from 
the fact that the three-brane carries Ramond charge is related to $f$ 
by \be (\pa C)_{\mu\nu\gamma\rho i} = \epsilon_{\mu\nu 
\gamma\rho}\pa_{i}f^{-1},\quad (\pa C)_{ijklm} = 
\epsilon_{ijklmn}\pa_{n}f .\ee
Note that $R^{4}= N/2\pi^{ 2}T_{3},$
where $N$ is the Ramond charge of the three-brane, and $T_{3}$ is the 
tension of the three-brane.

The world-volume action of a test three-brane in the background of 
such an extremal three-brane is given by the sum of a geometric 
induced volume term and a term arising {}from the coupling to the 
background 4-form field. The volume term and the 4-form term appear 
in precisely the manner needed for a vanishing static potential for 
the test three-brane. In static gauge, the action, for motions in 
which the test three-brane stays parallel to the three-brane 
producing the background, is then
\be S_{test}= \int d^{4}x \left[{T_{3}\over 2} v^{2} + {N\over 
16\pi^{2}} {v^{4}\over r^{4}} + \ldots \right]\ . \ee We want to 
compare this world-volume action to the Yang-Mills theory description.
{}From the effective actions computed in the first part of this 
paper, it is trivial to see that, with $A^{i} \equiv \phi^{i} \equiv 
g\sqrt T_{3} x^{i},\ i\in \{4,\ldots,9\},$ we get precise agreement 
between SYM theory and supergravity for the background field 
calculation (eq.~\ref{below}),
but there are no terms in the supergravity test particle 
action that correspond to four derivative terms with fewer than four 
scalar fields that {\it must} arise in the calculation with off-shell 
supersymmetry, eq.~\ref{above}. 

What is the physical meaning of this discrepancy? The $S$-matrix is, 
in general, related to the effective action evaluated at a stationary 
point (with prescribed asymptotics) of the effective action. The 
background-field calculation has only on-shell supersymmetry, hence 
the effective action computed {}from it possesses only supersymmetry 
using the {\it tree-level} equations of motion.  There is no 
reason why radiative corrections in 
effective actions in formalisms with or without off-shell
supersymmetry should agree, except when using tree-level equations of 
motion.  This is entirely 
consistent with the two calculations given above, since the 
tree-level equations of motion indeed set higher derivative terms to 
zero, and therefore lead to an agreement between the two 
calculations. However, since we are interested in the $S$-matrix, we 
do not want to use the tree-level equations of motion, but rather the 
loop-corrected equations of motion. Then, it would seem that the 
effective action computed in a manner preserving off-shell 
supersymmetry is the correct action to use for computing the $S$-matrix.
Carrying this line of thought to its logical conclusion, it would 
seem that one must use an off-shell formulation that preserves {\it 
all} of the supersymmetries to get definitive answers. 

Another way to state the 
key point here is that the $v^{4}/r^{4}$ potential term appears only at 
one-loop order.  Scattering solutions are therefore analogous to 
solitons in theories where symmetry breaking only occurs due to
radiative corrections, and care is needed in applying perturbative
recipes\cite{eweinberg}.   Effective actions with different 
numbers of 
off-shell supersymmetries need not agree for  configurations
that break some of the supersymmetry, such as those with non-zero 
velocities.
For long distance phenomena,  the equations of motion give
\be a \sim g^{2} {v^{4}\over r^{5} }\ee
and hence
\be {v^{2}a\over r^{3}} \sim g^{2} {v^{6}\over r^{8}},\ee 
which is a 
two-loop effect. The agreement  with supergravity at the lowest order 
is 
therefore unaffected.  {\it If} there had been a tree level potential,
this would not be the case.

Since the two-loop term, calculated in the background field formalism, 
has been shown to agree between supergravity and Matrix theory for 
D0-branes without acceleration\cite{bbpt}, it is of some interest 
to check if an off-shell calculation in the D0-brane case is 
consistent with this check. A na\"\i ve computation suggests that a 
correspondence between D0-brane and D3-brane terms of the following 
form
\be {1\over r^{7}} \leftrightarrow {1\over 15\pi^{2}r^{4}}, \qquad 
{1\over r^{6}} \leftrightarrow {{1\over 32\pi r^{3}}}, \qquad {1\over 
r^{5}} \leftrightarrow {1\over 6\pi^{2}r^{2}}. \ee Thus, inclusion of 
a term of the form $v^{2}a/r^{6}$ at one-loop order in
the D0-brane calculation would lead to a  contribution at 
two-loop order of the
form $(105/4\pi) v^{6}/r^{14}.$ This {\it additional} contribution 
would imply a {\it disagreement} between supergravity and D0-brane 
dynamics at two-loop order.
A check of this  term directly in D0-brane quantum 
mechanics is under way.

\section{Acknowledgements}
We are grateful to Diego Bellisai, Marc Grisaru, Igor Klebanov, Gilad 
Lifshytz,
Yaron Oz, Martin
Ro\v{c}ek, Yuri Shirman, and Wati Taylor for useful discussions. 
This work was supported in part by NSF grant PHY96-00258.

\newcommand{\NPB}[1]{{\sl Nucl. Phys.} {\bf B#1}} 
\newcommand{\PLB}[1]{{\sl Phys. Lett.} {\bf B#1}} 
\newcommand{\PRL}[1]{{\sl Phys. Rev. Lett.} {\bf B#1}} 
\newcommand{\PRD}[1]{{\sl Phys. Rev.} {\bf D#1}} 
\newcommand{\xxx}[1]{\mbox{hep-th/{#1}}}

\end{document}